\documentclass[12pt,showpacs,amsmath,amssymb,aps]{revtex4}
\usepackage{graphicx}
\usepackage{dcolumn}
\usepackage{bm}
\begin{document}
\title{Wang-Landau Monte Carlo simulation of
isotropic-nematic transition in liquid crystals}
\author{D. Jayasri${}^{\star}$, V. S. S. Sastry${}^{\star}$ and
K. P. N. Murthy${}^{\dagger}$}
\affiliation{${}^{\star}$School of Physics, University of Hyderabad,
                 Hyderabad 500 046 Andhra Pradesh, India\\
        ${}^{\dagger}$ Materials Science Division,
           Indira Gandhi Centre for Atomic Research,\\
         Kalpakkam 603 102, Tamilnadu, India}
\date{18 March 2005}
\begin{abstract}
Wang and Landau  proposed recently, a simple  and flexible
non-Boltzmann Monte Carlo method for estimating the density
of states, from which the macroscopic properties of
a closed  system can be calculated.
They demonstrated their algorithm by considering systems 
with discrete energy spectrum. We
find that the Wang-Landau algorithm does not perform well 
when the system has
continuous energy spectrum. We propose in this paper modifications
to the algorithm and demonstrate their 
performance 
on a lattice model
of liquid crystalline system (with Lebwohl-Lasher interaction having continuously 
varying energy),
exhibiting transition from high
temperature isotropic  to low temperature nematic phase.
\end{abstract}
\maketitle
\begin{section}{Introduction}
Monte Carlo methods have  emerged as a powerful and reliable tool
for  simulating several complex phenomena in statistical physics, 
see {\it e.g.} \cite{DPLKB,KPN}.
The  Metropolis algorithm
\cite{Metropolis} discovered in the middle of the last century can be 
considered as the starting point.  This algorithm
generates  a Markov chain, the asymptotic part of which 
contains microstates belonging to a
canonical ensemble at a temperature chosen for the simulation.
Expectation value of a macroscopic property can be estimated by taking a simple
arithmetic average over a Monte Carlo sample. The associated
statistical error is inversely proportional to the square root of
the sample size. Thus, in principle, we can estimate a physical
property to the desired accuracy by simply increasing the sample
size. However, if successive microstates in the sampled Markov chain
are correlated, the statistical error increases by a factor
$\sqrt{1+2\tau^{\star}}$, see {\it e.g.} \cite{Muller_Krumbhaar},
where $\tau^{\star}$ is the integrated correlation time. Such a
situation obtains when we simulate a system close to  criticality.

The Metropolis algorithm and its several variants like
Glauber \cite{glauber}, heat-bath \cite{cruetz}
and Kawasaki exchange \cite{kawasaki} algorithms
come under the
class of Boltzmann sampling techniques. The limitations of Boltzmann
sampling have long since been recognized. For example it can not
address satisfactorily problems of critical slowing down, {\it i.e.}
divergence of $\tau^{\star}$ with increase of system
size, near continuous phase transition. Cluster algorithms
\cite{cluster_algorithms} overcome this problem. Boltzmann sampling
is also not suitable for problems of super critical slowing down
near first order phase transitions. The microstates representing the
interface between  ordered and disordered phases have
intrinsically low probability of occurrence in a closed system and
hence are scarcely sampled; switching from one phase to the other
takes a very long time due to the presence of high energy barriers
when the system size is large; as a result the relative free
energies of ordered and disordered phases can not be easily and
accurately determined. 
Finally it is quite difficult to estimate
the absolute values of entropy or free energies in Boltzmann sampling
techniques.

\subsection{Non-Boltzmann sampling}
That non-Boltzmann sampling can provide a legitimate and often superior 
alternative to
Boltzmann sampling was  recognized even during the early days of Monte
Carlo practice, see {\it e.g.} \cite{Fosdick}. However, the practical
convenience and significance of non-Boltzmann sampling 
was appreciated only in the
middle of seventies when Torrie and Valleau \cite{TV} proposed the
so called umbrella sampling; this is  a fore-runner to all the
subsequent non-Boltzmann sampling techniques including the
multicanonical Monte Carlo \cite{Berg} and its several and recent
variants. Entropic sampling \cite{Lee}, equivalent to multicanonical
sampling \cite{Berg}, provides a transparent and 
intuitively appealing insight
into non-Boltzmann Monte Carlo techniques. 
It is based on the following premise.

The probability that a closed system can be found in a microstate
${\cal C}$ is given by
\begin{eqnarray}
P( {\cal C}) & = & \Big[ Z(\beta)\Big]^{-1}\exp\Big[-\beta E( {\cal
C})\Big].
\end{eqnarray}
In the above $E({\cal C})$ is the energy of the microstate ${\cal C}$ 
and  $\beta=(k_B T)^{-1}$, where $k_B$ is the Boltzmann constant and  $T$  is the temperature.
$Z(\beta)$ is the canonical partition
function given by,
\begin{eqnarray}
Z(\beta) & = & \sum_{ {\cal C}}\exp\Big[-\beta E( {\cal C})\Big]\nonumber\\
         & = & \int D(E)\exp(-\beta E)dE,\nonumber
\end{eqnarray}
where $D(E)$ is the density of states. 
The probability density for a
closed system to have  an energy $E$ is given by,
\begin{eqnarray}
P_B(E)\propto D(E)\exp(-\beta E)
\end{eqnarray}
where the suffix $B$ indicates that it is the Boltzmann-Gibbs  distribution 
(appropriate for modeling a closed system). 
Let us suppose  we want to sample microstates
in such a way that the resultant probability density of energy is
given by, 
\begin{eqnarray}
P_g (E)\propto D(E)\Big[g(E)\Big]^{-1}\label{g_ensemble},
\end{eqnarray}
where $g(E)$ is chosen as per the desired non-Boltzmann distribution.
Once $P_g (E)$ is known, an ensemble consistent with Eq. (\ref{g_ensemble})
can be constructed as follows. 

Let ${\rm C}_i$ be the current microstate
and ${\rm C}_t$ the trial microstate obtained from ${\rm C}_i$ by
making a local change, {\it e.g.} flip a randomly chosen spin in
Ising model simulation. Let $E_i=E({\rm C}_i)$ and $E_t=E( {\rm
C}_t)$ denote the energy of the current and of the trial microstate,
respectively. The next  entry  ${\rm C}_{i+1}$ in the Markov chain
is taken  as,
\begin{eqnarray}
{\rm C}_{i+1} = \begin{cases} {\rm C}_t\ {\rm with\ probability}\ $p$,\\
                                    {}\\
                        {\rm C}_i\ {\rm with\ probability}\ (1-p),
                        \end{cases}
\end{eqnarray}
where the acceptance probability $p$ is given by,
\begin{eqnarray}
p={\rm min}\left[ 1,\ \frac{P_g(E_t)}{p_g(E_i)}\right]
\equiv {\rm min}\left[ 1,\ \frac{g(E_i)}{g(E_t)}\right].
\end{eqnarray}
We call this non-Boltzmann sampling with respect to 
a given $g(E)$. 
It is easily verified that the above acceptance rule obeys
detailed balance. Hence  the
Markov chain constructed  would converge asymptotically
to the desired $g$-ensemble. 

When $[g(E)]^{-1} = \exp(-\beta E)$ we
recover   conventional Boltzmann sampling, implemented in the
Metropolis algorithm. For any other choice of $g(E)$
we get the corresponding non-Boltzmann sampling. 
Now, canonical ensemble average of
a macroscopic property $O( {\rm C})$  can be obtained by
un-weighting and re-weighting of $O( {\rm C})$ for each ${\rm C}$
sampled from the $g$-ensemble. For un-weighting we
divide by $[g( E( {\rm C}))]^{-1}$ and for re-weighting we
multiply by $\exp[-\beta E( {\rm C})]$. The weight factor associated with 
a microstate ${\rm C}$ belonging to the $g$-ensemble is thus,
\begin{eqnarray}
W({\rm C},\beta) &=& g\Big( E({\rm C})\Big)\exp\Big[ -\beta E({\rm C})\Big].
\label{weight_g}
\end{eqnarray}
We then have,
\begin{eqnarray}
\left\langle O\right\rangle &=&\frac{\sum_{ {\rm C}}
O( {\rm C}) W({\rm C},\beta)  }
{\sum_{ {\rm C}}W({\rm C},\beta)}.
\label{re_weighting}
\end{eqnarray}
The left hand side  of the above is the equilibrium value of $O$ in
a closed system at $\beta$, while on the right side the summation in
the numerator and in the denominator runs over microstates
belonging to the non-Boltzmann $g$-ensemble. It is also clear
that from a single simulation of a $g$-ensemble, we can calculate 
the canonical average of $O$ at various temperatures.
\subsection{Entropic sampling}
Entropic sampling obtains when $g(E)=D(E)$.
This choice  of $g(E)$ renders $P_g(E)$ the same for all $E$,
see Eq.~(\ref{g_ensemble}). The system does a simple
random walk on a one dimensional energy space.
Hence all energy regions are visited with equal probability. 
As a result, in the case of first order phase transition for example,
the
microstates on the paths (in the configurational space)
that connect ordered and disordered
phases  would  get
equally sampled. A crucial issue that remains to be clarified
pertains to the observation that we do not know
$D(E)$  $\grave{a}$ {\it priori}.

In entropic sampling we employ a strategy to push   $g(E)$ closer
and closer to $D(E)$, iteratively. We divide the range of energy into
a  large number of bins of equal widths. We denote the discrete-energy
version of $g(E)$ by the symbol $\{ g_i : i=1,\ 2,\ \cdots \}$. We start with 
$ \{ g_i^{ (0)}=1  \ \forall \ i \}$;  the superscript
is iteration run index and the subscript is  energy bin index. The
aim is to update $\{ g_i\}$ from one iteration to the next: $\{ g_i ^{
(0)}\}\to \{ g_i ^{ (1)}\}\to \cdots \{ g_i ^{ (k)}\}\to \cdots$, so
that asymptotically we get $\{ g_i\}$ as close to $\{ D_i\}$ as
desired, where $\{ D_i\}$ is the discrete energy representation of
$D(E)$. The iteration is carried out as follows. In the $k-th$
iteration, for example,  we generate a large number of microstates
employing acceptance probability based on $\{ g_i ^{(k)}\}$ and accumulate
an histogram $\{ h_i : i=1,\  9L^3\}$ of energy 
of visited microstates. 
We update $\{ g_i ^{(k)}:i=1,\ 9L^3\}$ to $\{
g_i ^{(k+1)}:i=1,\ 9L^3\}$, as given below,
\begin{eqnarray}
g_i ^{ (k+1)} = \begin{cases} g_i ^{(k)}\ & {\rm if}\ h_i = 0,\ \  \\
                           {}\\
                     g_i ^{(k)}\times h_i \ & {\rm if}\ h_i \ne\ 0 ,
                \end{cases}
\end{eqnarray}
for all $i=1,\ 2,\ \cdots ,\ 9L^3$. The updated $\{ g_i ^{ (k+1) }\}$ is employed in the next {\it i.e.}
$(k+1)-th$ run, during which a fresh histogram of energy is generated. 
After each run, the histogram is examined for its
uniformity. Flatter the histogram, closer is $\{ g_i \}$ to $\{ D_i
\}$. Thus, the calculated histogram serves two purposes
in entropic sampling, one for updating $\{ g_i \}$ and the other for
monitoring the convergence of $\{ g_i \}$ to $\{ D_i \}$. However, 
it is often neither practical nor necessary to get a strictly
flat histogram; an approximately flat histogram would be adequate,
thanks to the un-weighting followed by reweighting with the Boltzmann rule 
while calculating the averages,  see Eqs. (\ref{weight_g},\ 
\ref{re_weighting}).
Hence  the calculated macroscopic properties would come out right,
even if $\{ g_i \}$ does not converge strictly to $\{ D_i \}$.

\subsection{Wang-Landau algorithm}
A simple and flexible variant to entropic sampling was proposed
recently by Wang and Landau \cite{WL}. The distinguishing feature of
this  algorithm is the dynamic evolution of the acceptance
probability, $p$; we update $\{ g_i \}$ after every Monte
Carlo step. Let us say the system visits a microstate in a Monte
Carlo step and let the energy of the visited microstate fall in
the $m$-th energy bin;  then $g_{m}$ is updated to $f\times g_{m}$,
where $f$ is the Wang-Landau factor, see below. The updated $\{ g_i
\}$ becomes operative immediately for determining the
acceptance/rejection criteria of the very next trial microstate. We
set $f=f_0$ for the zeroth run; $f_0$ can be any number greater
than unity; the choice of $f_0=e$ has been originally
recommended by Wang
and Landau. We generate a large
number of microstates employing the dynamically evolving $p$. At the
end of a run we calculate the histogram of energy of microstates
visited by the system during the run. Because of the continuous updating
of $p$, the energy 
span of the density of states increases significantly and  
the energy histogram  serves 
to monitor the convergence of $\{ g_i \}$ to $\{ D_i \}$.
A run should be long enough to facilitate the system to
span  the energy over the  desired range  and to render the histogram
of energy approximately flat. At the end of,  say,  the $\nu$-th run,
the Wang-Landau factor for the next run is set as $f=f_{\nu+1} =
\sqrt{f_{\nu}}$. After several runs, this factor 
would be very close to unity; this implies that there would occur no
significant change in $\{ g_i\}$ during later runs. For example with
the square-root rule 
and $f_0=e$, we have
$f_{25}=\exp(2^{-25})\lesssim1+10^{-7}$. It is clear that $f$ decreases
monotonically with increase of the run index and 
reaches unity asymptotically. Wang and Landau have recommended the square-root
rule \cite{WL}; any other rule consistent with the above properties  of
monotonicity and asymptotic convergence to unity should do equally
well.

From the converged
$g$ 
the desired macroscopic properties of the system can be calculated; to this
end we invoke the
the connection between the density of states and
microcanonical entropy,
$\alpha(E)=k_B\log D(E)$. Thus the Monte Carlo estimate of
microcanonical entropy is $k_B\log g(E)$.
For implementing such a scheme we need to
normalize $g(E)$. The normalization
constant  should be obtained from known
properties  of the system. For example 
in Ising model, the ground state is doubly
degenerate: $D(E_{{\rm min}})=2$.  The total number of microstates equals $2^V$
where $V$ is the number of Ising spins in the Monte Carlo model: 
$\int_{E_{{\rm min}}}^{E_{{\rm max}}}D(E)dE=2^V$.
Either of these known information can be employed
for normalizing $g$. The normalized $g(E)$ provides a
good approximation to $D(E)$. 

Alternately,  we can take the output $\{ g_i\}$ 
from the above  and carry out a single
long non-Boltzmann sampling  run which generates microstates 
belonging to
the $g$-ensemble. (Note that during the production run
we do not update $g(E)$.
By un-weighting and re-weighting 
of the microstates generated in the production
run, we calculate the desired properties of the system as a function
of $\beta$. This is the strategy we shall follow for the 
simulation of liquid crystal system, described  in the rest of the   paper. In
this strategy, we can employ arbitrary normalization of $g$; more
importantly, it is adequate  if 
\begin{enumerate}
\item[(a)] 
the system visits the energy
region  of interest and  not necessarily the entire range and 
\item[(b)]
the histogram of energy in the region of interest is approximately flat.
\end{enumerate}

The usefulness of the Wang-Landau algorithm has been unambiguously
demonstrated for systems with discrete energy spectrum. However,
when we try to apply this technique
to systems with continuous energy, there are serious difficulties.
Liquid crystalline materials
with continuous energy spectrum provide such an example.
In this paper  we report  simulation of  a 
liquid crystalline system focusing 
attention on nematic-isotropic transition.
The paper is organized as  follows. In section II,
we describe a lattice model and Hamiltonian  of a liquid
crystal system. We simulated the system with the conventional Wang-Landau
algorithm.
We find that the
dynamics becomes extremely slow even for moderately large  systems. The 
system gets stuck in certain regions of the configurational space.
This problem
appears to be generic to the algorithm when applied to continuous
energy systems.  Hence we modify the Wang-Landau algorithm and 
the details of the simulation are given in 
section III. The results on temperature variation of 
various macroscopic properties of the system are discussed in section IV. In section V 
we briefly summarize the work and highlight its salient features.  
\end{section}

\begin{section}{Lattice model of bulk liquid crystals}
We consider an $L\times L\times L$ cubic lattice
with each lattice site holding a three dimensional unit
vector $\vert u\rangle$, called a spin. 
The elements of the vector are the direction
cosines of a 'spin'  in a laboratory frame of reference.
A 'spin'  represents notionally, a
single uniaxial liquid crystal molecule
or more realistically a  cluster  
containing typically a hundred of them. 
The spins are actually
'headless'  in the sense the system has 
head-tail flip symmetry.
Two nearest neighbour spins interact with each
other as per a potential  proposed by Lebwohl and
Lasher (LL) \cite{LL} which has such a  head-tail flip symmetry.
The interaction energy is given by,
\begin{eqnarray}
  \epsilon_{i,j} = -\frac{1}{2}
  \left[ 3\cos ^2 (\theta _{i,j}) -
   1\right],\label{eq_LL_potential}
\end{eqnarray}
where $i$ and $j$ are nearest neighbour lattice sites.
$\theta_{i,j}$ is the angle between the
two spins: $\cos (\theta_{i,j})=
\left\langle u_i\vert u_j\right\rangle$.
The interaction energy of a single
nearest neighbour pair of spins
ranges from a minimum of $-1$,
when $\theta_{i,j}=0$, or equivalently $\langle u_i\vert u_j\rangle=1$, 
to a maximum of
$+1/2$, when $\theta_{i,j}=\pi/2$  or equivalently 
$\langle u_i\vert  u_j\rangle=0$.
Total energy of the system  in
microstate ${\cal C}$ is given by,
\begin{eqnarray}
 E({\cal C})=\sum_{\left\langle i,j\right\rangle}
\epsilon_{i,j},
\end{eqnarray}
where the sum runs over all distinct nearest
neighbour pairs of lattice sites in the system taking
into account the periodic boundary conditions in all the
three directions. The total energy of the system thus varies
continuously from a minimum of $-3L^3$ to a maximum of $+3L^3/2$.
When the system is completely ordered with all the spins aligned,
the energy is minimum and equals $-3L^3$; the energy is zero 
for an isotropic (completely disordered) phase. 
We calculate several macroscopic properties
of the liquid crystalline system
and report here  results which include 
orientational order
parameter $\langle S\rangle $, average energy 
$\langle E\rangle$,
specific heat (at constant volume) $C_V$, 
and the Binder's
reduced fourth cumulant of energy $V_4$.
\end{section}

First we employed conventional Wang-Landau algorithm and carried out Monte 
Carlo simulation of the lattice model of the Liquid crystalline system.  
We found that the dynamics was extremely slow when the system  size $L$ 
is $6$ and above. The calculated density of states $g(E)$ becomes 
steeper with increase of Monte Carlo iterations.
As a result the system gets
stuck in certain narrow regions.
There is practically no evolution of the calculated 
density of states $g(E)$. Increasing the number of Monte Carlo steps 
in a Wang-Landau
iteration  does not seem to remedy the situation. Instead, sharp peaks 
emerge and grow at either ends 
of $g(E)$. These problems appear to be generic to the Wang-Landau 
algorithm when applied to continuous energy systems. To overcome 
them, we experimented with several 
modifications \cite{JSM} of the 
algorithm and finally arrived  at  
a strategy described 
in the next section.
\begin{section}{Modified Wang-Landau Monte Carlo simulation
                of bulk liquid crystal system}
Free wheeling spins are placed on the vertices  of a three 
dimensional cubic lattice with their orientations sampled 
randomly and independently. Orientation of a spin is 
specified by the polar angle $\theta$ and an azimuthal angle 
$\phi$ with respect to a laboratory fixed three dimensional 
co-ordinate system. We sample $\mu=\cos(\theta)$ from a
uniform distribution between $0$ and $1$, and the azimuthal 
angle $\phi$ from a uniform distribution between $0$ and 
$2\pi$. We divide the energy range $(-3L^3,+1.5L^3)$
into $9L^3$ number bins of equal widths $\Delta E=0.5$. 
We start with an array $\{ g_i =e^2\ \forall\ i=1,\ 9L^3\}$.
Let ${\rm C}_0$ denote the initial microstate constructed by 
placing spins with random and independent orientations at the 
lattice sites. Let $E( {\rm C}_0)$ fall in 
the $\mu -th$ energy bin.
We select randomly a lattice site and make a random change  
in the  orientation of the spin residing at  that site. We 
employ Barker's method \cite{Barker_Watts} to generate a 
new trial orientation from the  current microstate. 
Let ${\rm C}_t$ 
denote the trial microstate and let its energy belong to the
$\nu$-th bin. If $g_{\nu}\le g_{\mu}$, we accept the trial 
state and set ${\rm C}'_1={\rm C}_t$; if $g_{\nu} > g_{\mu}$, 
we calculate the ratio $p=g_{\mu}/g_{\nu}$. We select a 
random number $r$ (uniformly distributed between  $0$ to $1$);
if $r\le p$ we accept the trial microstate: ${\rm C}'_1={\rm C}_t$. 
Otherwise we reject ${\rm C}_t$ and set ${\rm C}'_1={\rm C}_0$.
This constitutes a single move. We continue in the same fashion and get,
${\rm C}_0\to{\rm C}^{\prime}_1\to {\rm C}^{\prime}_2\cdots{\rm C}
^{\prime}_{L^3 -1}\to{\rm C}_1$. A set of $L^3$ moves  constitutes a
Monte Carlo Sweep  (MCS). In the first MCS, since $g_i$  
is the same for all $i$, every move gets accepted. 

At the end of the  MCS, let us say  the system is in 
microstate ${\rm C}_1$. Let $E( {\rm C}_1)$ belong to the $k$-th
energy bin. We update $g_k$ to $f\times g_k$, where $f=f_0$.
The updated $\{ g_i \}$ becomes operative for deciding 
acceptance/rejection in the next $L^3$ moves  that constitute 
the next MCS leading to ${\rm C}_2$. Thus we get a chain of 
microstates ${\rm C}_0\to{\rm C}_1\to \cdots{\rm C}_N$. We 
take $N=10,000$. Generating a Markov chain of length
$N$ constitutes one  iteration. For the next   
iteration we change $f$ to $f^{0.9}$. The microstate generated 
at the end of an iteration forms the initial microstate 
for the next. Also the updating of the density of states is continued 
from one iteration to the next. We carry a total of $M$  
iterations and this constitutes a Wang-landau (W-L) run, 
see below. In the last
iteration of a W-L run, we have  
$f=f_M=f_0^{\mu(M)}$, where $\mu(M)=(0.9)^M$. We have chosen
$M=160$ so that $f_M - 1\approx 10^{-7}$ for $f_0=10$. 

We start a W-L run  with $f$ reset to $f_0$.  The 
microstate generated at the end of a W-L run 
is taken as the initial microstate for the next. 
Similarly the updated density of states $\{ g_i\}$ 
at the end of a W-L run
provides the initial density of states for the next. We carry out a 
total of $50$ W-L runs. The value of $f_0$ is $100$ for the first 
forty, $10$ for the next $9$ and $f_0=e$ for the 
last W-L run. 

The density of states at the end of 
$50$ W-L runs  is taken as an input for  a long 
non-Boltzmann sampling 
run of $2.5$ million Monte Carlo sweeps, called 
the production run.
Thus we get a $g-$ensemble of 
microstates from which the desired macroscopic properties can be 
calculated by un-weighting and re-weighting. 

We also found that it is imperative to  employ numerical 
techniques that avoid overflow problems and the attendant loss of precision 
due to truncation. To this end,
we adapted the techniques suggested by Berg \cite{BAB}.
These involve principally the following. Let $\alpha_i=\log g_i$ denote the 
microcanonical entropy. We define $\xi_i=\log \alpha_i$. 
We carry out all the calculations
in terms of $\{  \xi_i : i=1,\ 2,\ \cdots,\  9L^3 \}$. 
We derive expressions for acceptance 
probability $p$ in terms of $\{ \xi_i\}$ 
and employ them in the simulation. Similarly 
we derive expressions for the updating of $\{ \xi_i\}$ and for 
un-weighting and re-weighting, in terms of 
$\{ \xi_i \}$. 
These are
briefly described in the appendix.
Employing this modified Wang-Landau algorithm we simulated 
a lattice model of liquid crystalline system with $L=4, 6, 8, 10 $ and $12$
focusing on nematic-isotropic transition.
We present the results in the next section.
\end{section}

\begin{center}
\begin{figure}[htbp]
\includegraphics[height=15cm,width=15cm]{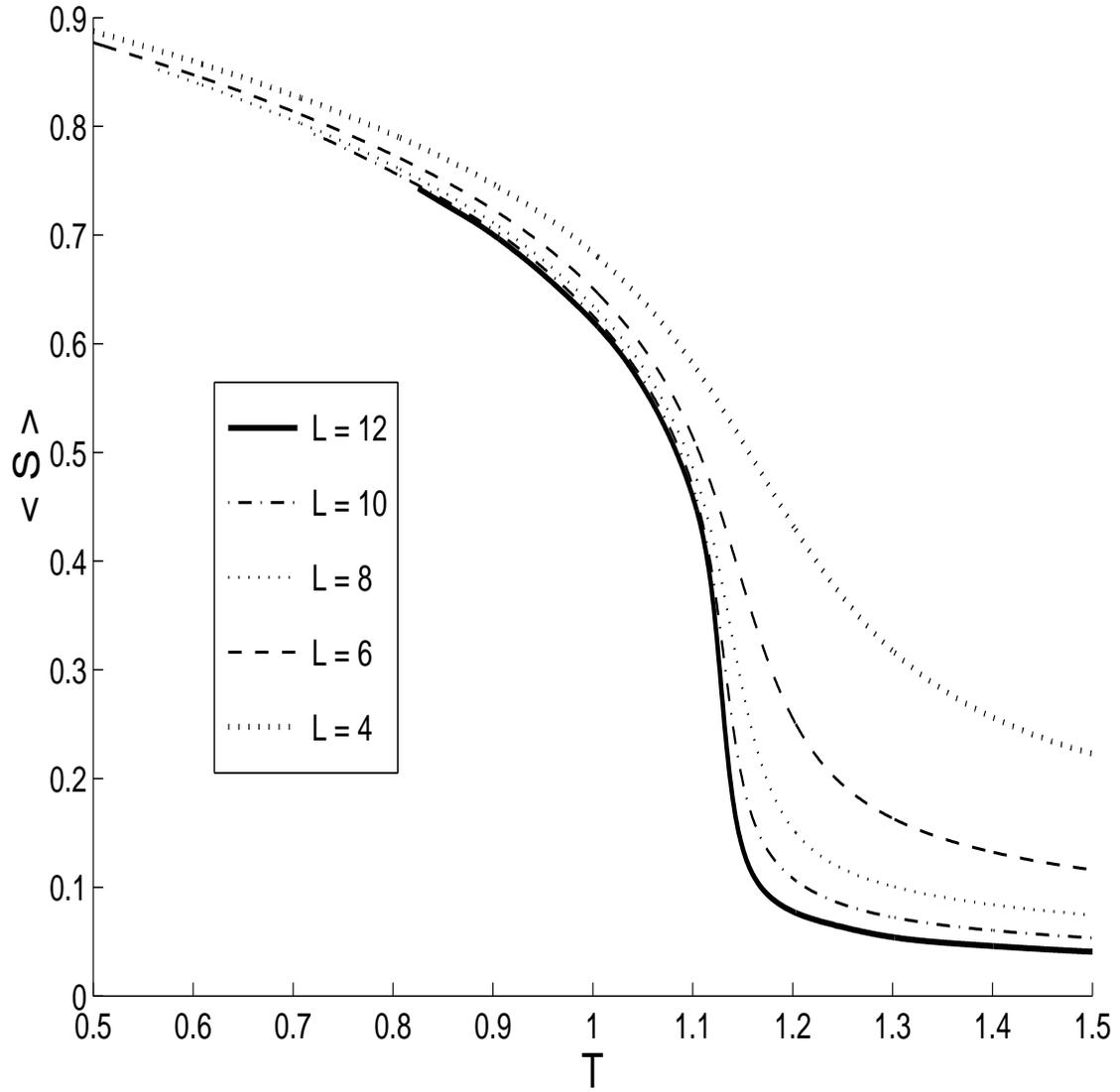}
\caption{The orientational order parameter $S$ {\it versus} temperature
for $L=4,\ 6,\ 8$, $10$ and $12$. The transition becomes sharper with increase
of system size}\label{ordpar_fig}
\end{figure}
\end{center}

\begin{center}
\begin{figure}[htbp]
\includegraphics[height=15cm,width=15cm]{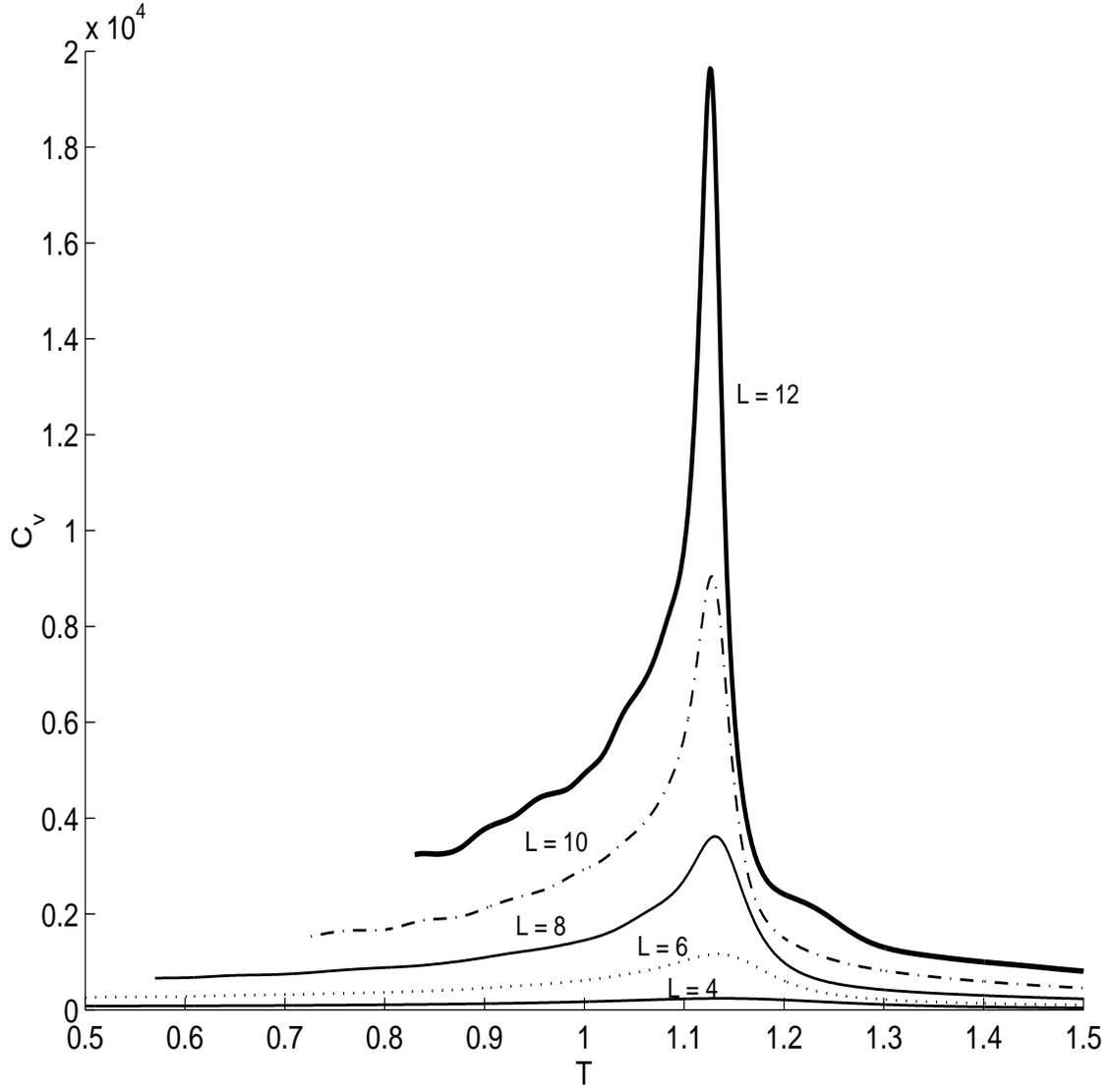}
\caption{Specific Heat $C_V$ as a function of
temperature for $L=4,\ 6,\ 8,\ 10,\ {\rm and} 12$; the transition becomes
sharper with increase of system size; the transition temperature  
(the value of $T$ at which the curve peaks) shifts to lower values with increase 
of system size.}\label{fig_Cv_ps}
\end{figure}
\end{center}

\begin{center}
\begin{figure}[htbp]
\includegraphics[height=15cm,width=15cm]{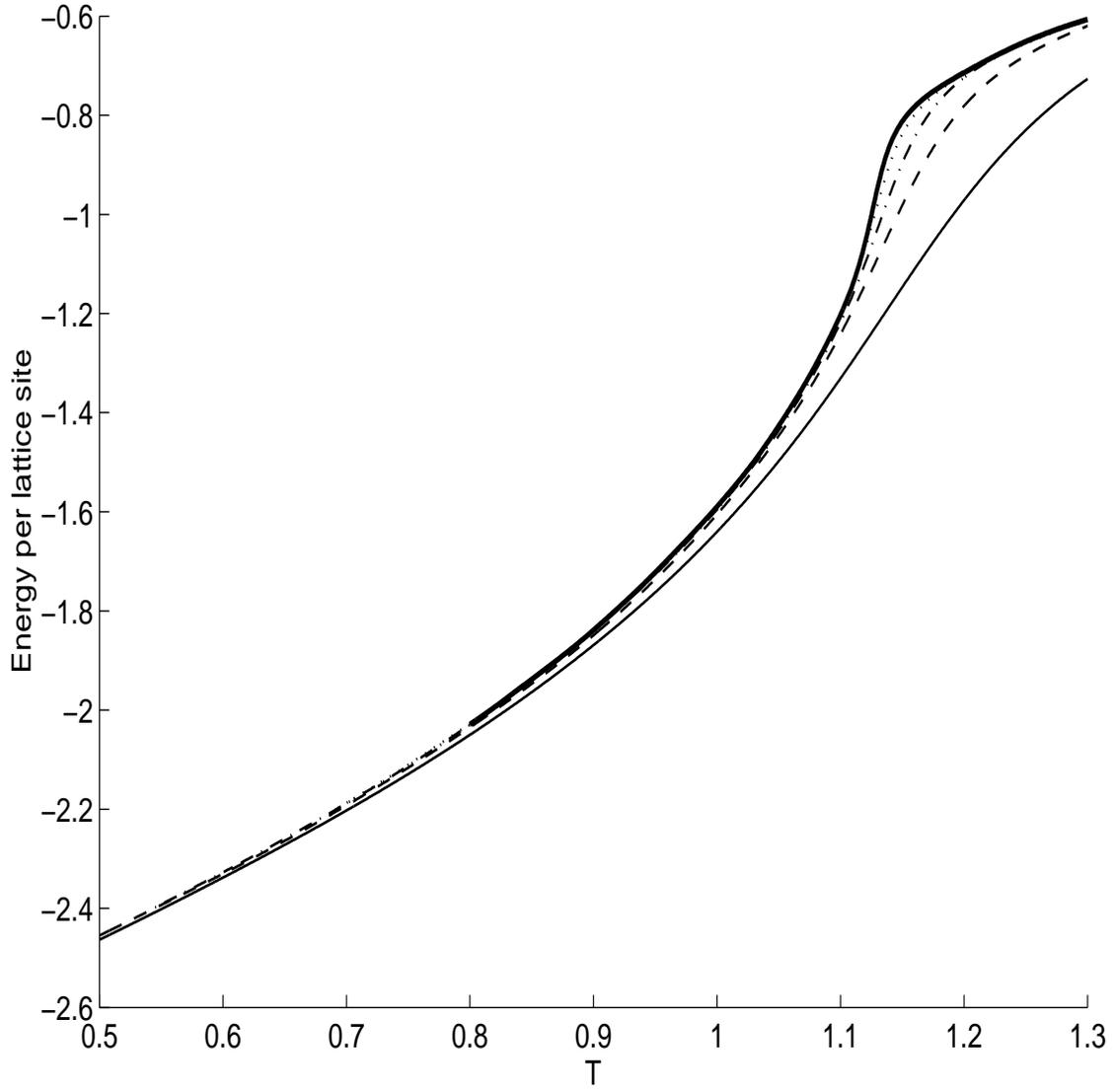}
\caption{Average energy   as a
function of temperature for $L=4,\ 6,\ 8,\ 10,\ {\rm and} 12$, 
from bottom to top respectively. The
transition becomes sharper with increase of system size}
\label{fig_energy}
\end{figure}
\end{center}

\begin{center}
\begin{figure}[htbp]
\includegraphics[height=15cm,width=15cm]{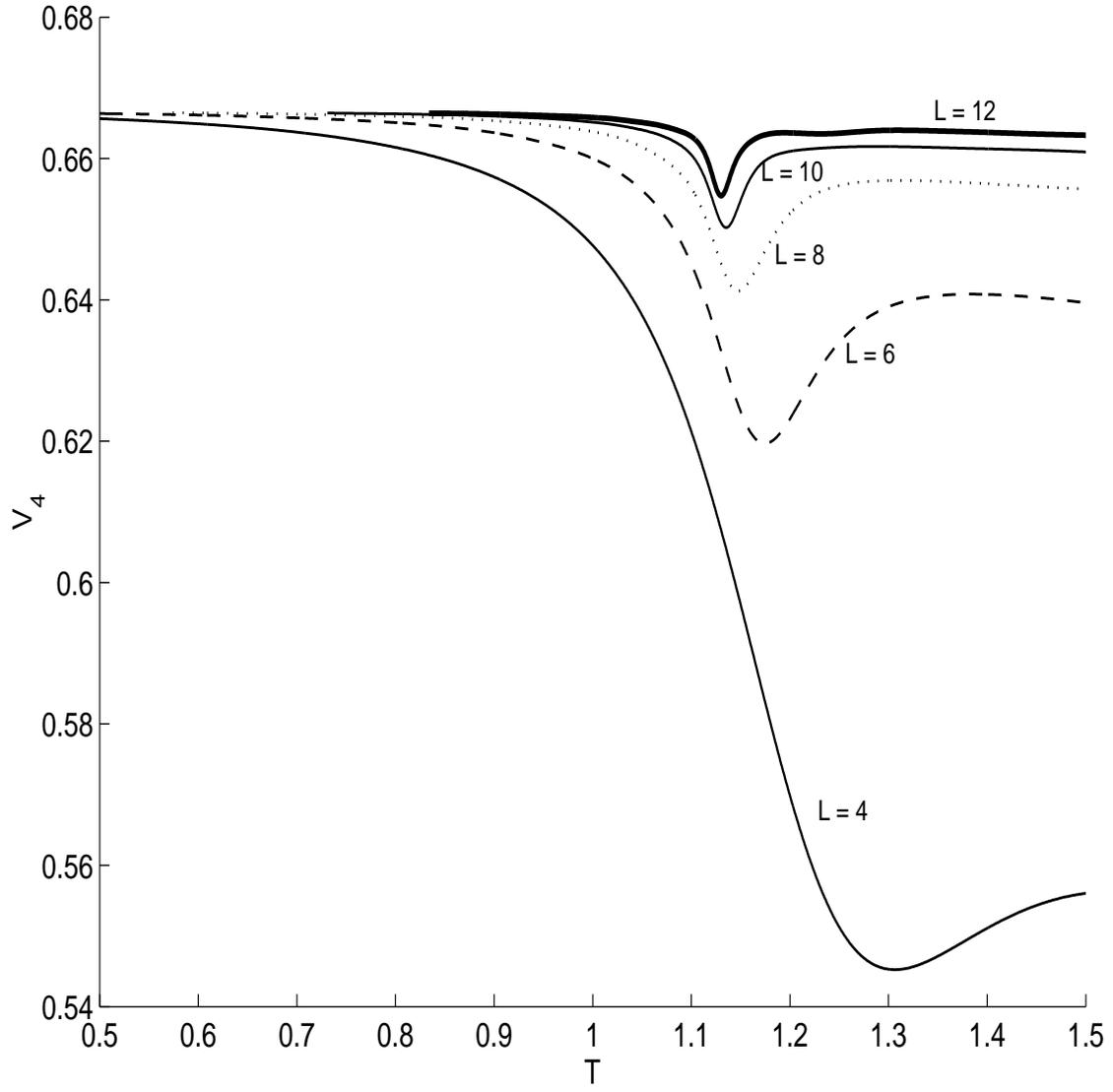}
\caption{Binder's fourth order cumulant of energy for $L=4,\ 6,\ 8,\
 10,\ {\rm and}\ 12$. The behaviour is indicative of first order
transition} \label{fig-BC}
\end{figure}
\end{center}

\begin{center}
\begin{figure}[htbp]
\includegraphics[height=15cm,width=15cm]{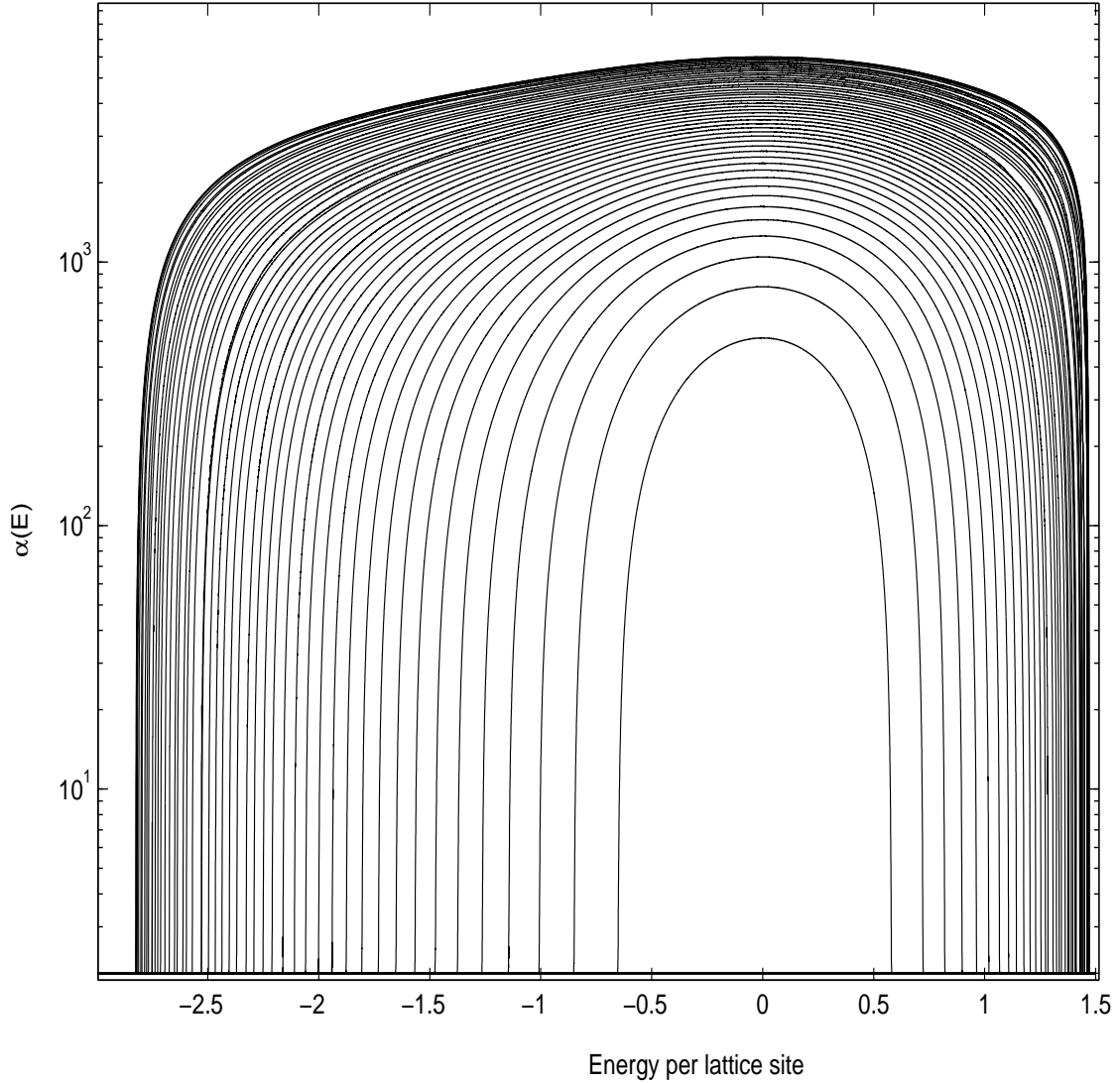}
\caption{Microcanonical entropy $\alpha_i=\log g_i$ as a function 
of energy at the end of successive outer iterations starting from the inner most curve to the 
outer most. The data correspond to $L=12$ and are plotted on a log-linear curve. 
The shape of the curve will correspond to that of $\chi$ employed in the simulation program.
The logarithm of the outermost curve is taken as the  input for a long production 
run which generates a $g-$ ensemble. Note that almost the 
entire energy range is spanned.} \label{log_log_g}
\end{figure}
\end{center}

\begin{center}
\begin{figure}[htbp]
\includegraphics[height=15cm,width=15cm]{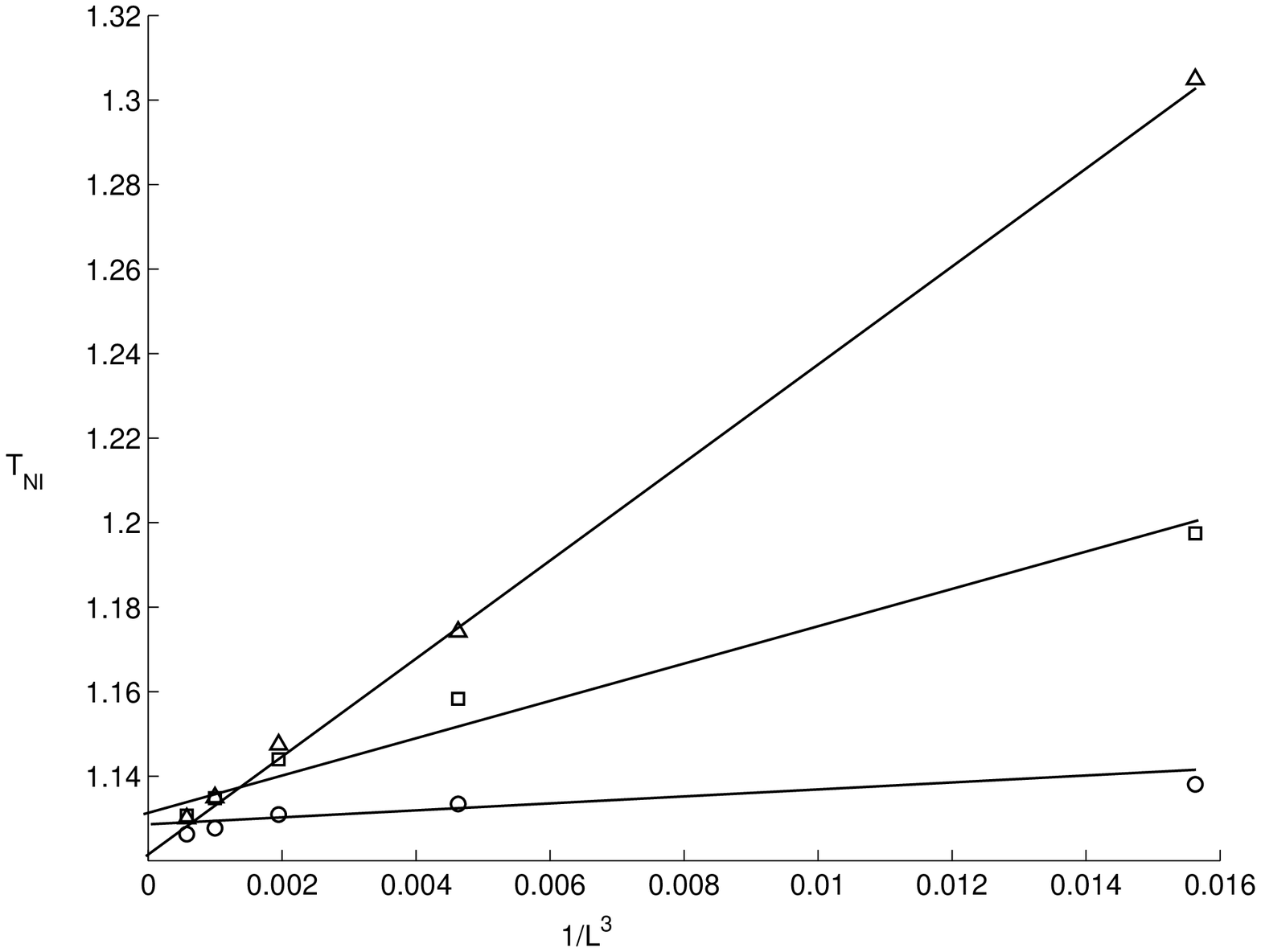}
\caption{The transition temperature versus $1/L^3$. The 
top line with symbol $\triangle$ corresponds to $T_{NI}$ obtained from  fourth 
cumulant of Binder; the middle line with symbol $\Box$ corresponds to
 $T_{NI}$ obtained from the orientational susceptibility and the bottom line 
corresponds to the  $T_{NI}$ obtained from the specific heat. The value of 
$T_{{\rm NI}}$ in the limit $L\to\infty$ is given by 
$1.1284$ from finite size scaling of specific heat data, $1.1299$ 
from data on susceptibility and $1.1211$ from data on Binder's fourth 
cumulant of energy.}
\label{scaling}
\end{figure}
\end{center}

\begin{section}{Characteristics of Bulk liquid crystals}
The orientational order parameter $S$
of a microstate is defined as follows.
Let $\vert u_i\rangle$ denote the spin
at the $i$-th lattice site. We first
construct an average projection operator
for a  microstate ${\rm C}$ given by,
\begin{eqnarray}
A({\rm C})=\frac{1}{L^3}\sum_{i=1}^{L^3} \vert u_i\rangle\langle u_i\vert\ .
\end{eqnarray}
From $A$  we construct a traceless symmetric tensor,
\begin{eqnarray}
Q({\rm C})=A-\frac{1}{3}{\rm trace} (A)\times I \ ,
\end{eqnarray}
where $I$ denotes
a unit matrix. Let $\lambda_{max}({\rm C})$ denote
the largest eigenvalue of $Q$.
Then $S({\rm C}) =3\lambda_{max}({\rm C})/2$.  The
corresponding eigenvector
$\vert\lambda_{{\rm max}}\rangle$
defines  the director for the microstate  ${\rm C}$.

Figure (\ref{ordpar_fig})  depicts $\langle S\rangle $ 
as a function of temperature  for system sizes $L=4,\ 6,\ 8,\ 10$ and $12$.
We observe that the modified Wang-Landau Monte Carlo simulation 
predicts correctly 
the  transition from a high temperature disordered (isotropic) phase to 
a low temperature nematic phase. For $L=4$ the transition is not sharp, 
due to finite size effects.
However when we increase the system size, 
the transition becomes sharper.

Next we investigate the behaviour of  
specific heat at constant volume $C_V$ as a function of 
temperature. 
$C_V$ is calculated
from  energy fluctuations and is given by,
\begin{eqnarray}
C_V = \frac{\left\langle E^2\right\rangle -
\left\langle E\right\rangle ^2}
           {k_B T^2}.
\end{eqnarray}
The results are depicted in  Fig. (\ref{fig_Cv_ps}).
As $L$ increases the $C_V$ profile becomes sharper. Also the 
temperature $T_{{\rm NI}}(L)$ at which the specific heat is maximum,
 shifts 
slightly to lower values, as expected. The transition temperature
for the LL model has been obtained earlier \cite{FB} and is
given by $T_{{\rm NI}}=1.1237\pm 0.0006$. We find
that $T_{{\rm NI}}(L=12)$ calculated from the $C_rV$ is 
$1.126$, in good agreement with the earlier estimate.
We can estimate the $L\to\infty$ limit of the transition temperature
by finite size scaling discussed later.

The variation of average energy $\langle E\rangle$ 
with temperature is depicted 
in Fig. (\ref{fig_energy}). This quantity decreases with decrease of 
temperature. At transition the fall is sharp for large $L$.

We have calculated  
Binder's reduced fourth order cumulant of energy  
denoted by the symbol $V_4$, see \cite{binder_cumulant}. 
It is given by,
\begin{eqnarray}
V_4 = 1- \frac{\left\langle E^4 \right\rangle}
              {3\left\langle E^2 \right\rangle^2}.
\end{eqnarray}
The variation of $V_4$ with $T$ is depicted in Fig. (\ref{fig-BC}) 
for $L=4,\ 6,\ 8,\ 10$ and $12$. 
Each curve 
shows a minimum at an effective  transition temperature. 
As the system size increases the effective transition 
temperature shifts to lower values
as expected. 
Also $V_4 \to  2/3$ for $T<<T_{{\rm NI}}$ for all $L$ considered  
and for $T>>T_{{\rm NI}}$ for large $L$. 
This is a 
a clear signature of a first order transition.  

Figure (\ref{log_log_g}) depicts microcanonical entropy $\alpha(=\log g)$ 
versus $E$ for system size $L=12$ on a log-linear graph. 
The curves depict the shape of $\xi$($\log \alpha)$ versus
$E$. The results on entropy after successive W-L 
runs are shown  starting from
the inner most and ending in the outer most. 
We see clearly that the range of energy spanned 
increases with increase of W-L runs. 
The outer most curve is the output of the last W-L run.. 
The data on $\{ \xi_i : i=1,9L^3\}$ obtained 
at the end of the last W-L run 
is  employed in the  long 
non-Boltzmann sampling run (production run) 
and a  $g$-ensemble of microstates is generated. 
All the quantities referred to above 
were calculated by un-weighting and re-weighting 
at temperatures spaced out with a fine 
resolution of $0.001$.

Finally we have presented in Fig. (\ref{scaling}) 
the finite size scaling of the transition temperature
obtained from specific heat, orientational susceptibility 
and the fourth order 
cumulant of Binder. 
The orientational susceptibility $\chi$ is calculated from the 
fluctuations of $S$ and is given by,
\begin{eqnarray}
\chi = \frac{ \langle S^2\rangle -\langle S\rangle^2}
            {k_B T}.
\end{eqnarray}
The transition temperature is plotted 
against inverse of the volume of the
system. The three curves scale linearly with $1/L^3$ 
and extrapolation  ($L\to\infty$) 
gives an estimate of the nematic-isotropic transition 
temperature. $T_{{\rm NI}}(L=\infty)$ estimated from specific heat data
is $1.1284$, from susceptibility data is $1.1299$ and from the data on 
Binder's cumulant is $1.1211$. These results are in good agreement with 
$T_{{\rm NI}}=1.126(5)$ - an earlier  estimate, see  \cite{FB} 
upto second decimal.
\end{section}
\begin{section}{conclusions}
We have demonstrated for the first time 
the applicability of the recently proposed
Wang-Landau Monte Carlo algorithm to the study of liquid
crystalline systems with continuous energy spectrum.
We have made use of the flexibility of the
algorithm and studied nematic-isotropic transition in a three
dimensional lattice model of bulk liquid crystals. We have employed
the Lebwohl-Lasher potential that has the  head-tail flip symmetry
of the nematic director and in which the energy varies continuously. 
For even moderately large system 
Wang-Landau dynamics becomes unacceptably slow. The density of states
function $g(E)$ gets confined to a narrow energy range and 
becomes steep.  
As a result, the system spans only a restricted range of
energy. Increasing the number of sweeps in an iteration  does not seem to 
help. This
slowing down of dynamics  seems to be an inherent problem of this
algorithm for such systems.  Interestingly such problems 
do not arise for simulating systems with discrete
energy spectrum {\it e.g.} Ising and Potts  spin models. 
To overcome these  problems, we have proposed a few 
modifications to the Wang-Landau algorithm and demonstrated that these
modifications help the  basic algorithm to  span larger regions of the energy
space systematically. We show that 
the macroscopic properties bulk liquid crystalline system
can be calculated with a good degree of accuracy and with vastly improved
temperature resolution. This opens up  the possibility of exploiting
the full power of the (non-Boltzmann) Wang-Landau 
Monte Carlo techniques to simulate several
complex phenomena in liquid crystalline systems. Examples of
such problems include phase transition 
in thin films deposited on substrates with complex geometry,
study of polymer dispersed liquid crystals (PDLC) and  effect
of disorder and confinement on the nematic-isotropic phase 
transition. Work on these 
and related problems are in progress and will be reported soon.
\end{section}
\section*{Acknowledgment}
The Monte Carlo simulations  reported in this paper were carried out
at the Centre for Modeling, Simulation and Design of the University 
of Hyderabad.  

\section*{Appendix}
Let $g_i$ denote the number of microstates in the i-th energy bin and
$\alpha_i=\log(g_i)$ the corresponding microcanonical entropy. We define 
$\xi_i=log(\alpha_i)$. In the program only the  array
$\{ \xi_i :  i=1,9L^3\}$ is stored, updated and eventually employed in
reweighting. All the required parameters like the acceptance probability $p$, and 
un-weighting cum re-weighting factor $W$ are calculated in terms of $\{ \xi_i :  i=1,9L^3\}$.

First we initialize $\{ \xi=\log(2)\ \forall\  i=1,9L^3\}$. Let the energy of the 
current microstate belong to an energy bin $c$ and the
trial microstate, to an energy bin $t$. The acceptance probability of
the trial state in the Wang-Landau algorithm is given by
by,
\begin{eqnarray}\label{p_xi}
p=\begin{cases}
1\ \ {\rm if}\ \ \xi_t\le \xi_c\\
                               \\
\exp
\bigg[
-\exp
\bigg\{
\xi_t +\log
\Big(
1-\exp
\left(
-\left(
\xi_t -\xi_c
\right)
\right)
\Big)
\bigg\}
\bigg]
\ \ {\rm if}\ \ \xi_c < \xi_t
\end{cases}\nonumber\end{eqnarray}

If the visited microstate  has an energy falling in the say $i$-th bin,
then $\xi_i$ is updated to $\xi_i+\log(\log(f))$ where $f$ is the 
Wang-Landau factor for that run. 

The un-weighting and re-weighting of 
microstates belonging to the $g-$ ensemble is carried out as follows.
Let ${\rm C}$ be the microstate under consideration. Let the energy $E=E({\rm C})$
of the microstate fall in the bin $c$. Let $\xi_c$ be the value of $\xi$ in that 
bin. The weight factor $W({\rm C})$ attached to ${\rm C}\in g-{\rm ensemble}$
is given by 
\begin{eqnarray}
W(C)=\begin{cases}
\exp\Bigg[ +\exp\bigg[ \ \ \ \xi_c \ \ +\ \ \  
\log\big\{ 1- \exp(-\Delta_1)\big\}\bigg]\Bigg] \ \ 
{\rm where}\  \Delta_1=\xi_c-\log(\beta E)\ge 0\cr
                                           \cr
\exp\Bigg[-\exp\bigg[ \log(\beta E) +\log\big\{ 1 - \exp(-\Delta_2)\big\}\bigg]\Bigg]
\ \ {\rm where}\  \Delta_2= \log(\beta E) - \xi_c \ge 0
\end{cases}\nonumber
\end{eqnarray}
It is easily verified that the above weight factor is 
identical to the one given by Eq. (\ref{weight_g}) except that we have expressed 
it in terms of $\{\xi_i\}$ instead of $\{g_i\}$. 
The average of a macroscopic property $O$ is calculated
employing Eq. (\ref{re_weighting}).

\newpage
\newpage
\newpage
\newpage
\newpage
\newpage

\begin{thebibliography}{100}
\bibitem{DPLKB}
D. P. Landau and K. Binder,
{\it A Guide to Monte Carlo Simulations in Statistical Physics},
Cambridge University Press (2000).
\bibitem{KPN}
K. P. N. Murthy, {\it Monte Carlo Methods in Statistical Physics},
Universities Press, India
(2004).
\bibitem{Metropolis}
N. Metropolis, A. W. Rosenbluth, M. N. Rosenbluth, A. H. Teller and E. Teller,,
J. Chem. Phys. {\bf 21}, 1087 (1953)
\bibitem{Muller_Krumbhaar}
H. M\"uller Krumbhaar and K. Binder,
J. Stat. Phys. {\bf 8}, 1 (1973)
\bibitem{glauber}
R. J. Glauber, J. Math. Phys. {\bf 4}, 294 (1979)
\bibitem{cruetz}
M. Creutz, Phys. Rev. Lett. {\bf 43}, 553 (1979)
\bibitem{kawasaki}
K. Kawasaki in {\it Phase transitions and critical phenomena},
vol. 2, Eds. C. Domb and M. S. Green, Academic, London (1972)
\bibitem{cluster_algorithms}
R. H. Swendsen and J.-S. Wang,
Phys. Rev. Lett. {\bf 58}, 86 (1987);
U. Wolff,
Phys. Rev. Lett. {\bf 62}, 361 (1989);
R. G. Edwards, and A. D. Sokal, 
Phys. Rev.,  D {\bf 38}, 2009 (1988)
\bibitem{Fosdick}
L. D. Fosdick, Methods Comput. Phys. {\bf 1}, 245 (1963)
\bibitem{TV}
G. Torrie and J. P. Valleau,
Chem. Phys. Lett. {\bf 28}, 578 (1974)
\bibitem{Berg}
B. A. Berg and T. Neuhaus,
Phys. Lett. B {\bf 267}, 249 (1991);
Phys. Rev. Lett. {\bf 68}, 9 (1992)
\bibitem{Lee}
J. Lee,
Phys. Rev. Lett. {\bf 71}, 211 (1993); Erratum, {\bf 71}, 2353 (1993)
\bibitem{WL}
F. Wang, and D. P. Landau, Phys. Rev. Lett. {\bf 86}, 2050 (2001);
Phys. Rev. E {\bf 64}, 056101 (2001).
\bibitem{LL}
P. A. Lebwohl and G. Lasher, Phys. Rev. {\bf A 6}, 436 (1972)
\bibitem{binder_cumulant}
M. S. S. Challa, D. P. Landau and K. Binder, Phys. Rev. B {\bf 34},
1841 (1986).
\bibitem{JSM}
D. Jayasri, V. S. S. Sastry and  K. P. N. Murthy,
{\it Wang-Landau Monte Carlo Simulation of Isotropic - Nematic
Transition under Confinement},
22$^{{\rm nd}}$ Int. Conf. Stat. Phys., 
         Bangalore, July 4-9, 2004.
\bibitem{Barker_Watts}
J. A. Barker and R. O. Watts, Chem. Phys. Lett. {\bf 3}, 144 (1969)
\bibitem{BAB}
B. A. Berg, {\it Multicanonical simulations step by step},
arXiv:cond-mat/0206333 18 June (2002).
\bibitem{FB}
U. Fabbri and C. Zannoni, Mol. Phys. {\bf 58}, 763 (1986)
\end{thebibliography}
\end{document}